\documentclass[11pt]{article}
\usepackage[margin=1.5in]{geometry}
\usepackage{amssymb}
\usepackage{graphicx}
\usepackage{cite}

\newcommand{\degC}{\ensuremath{^{\circ} \textrm{C}}}

\begin{document}

%\begin{frontmatter}

%% Title, authors and addresses

%% use the tnoteref command within \title for footnotes;
%% use the tnotetext command for theassociated footnote;
%% use the fnref command within \author or \address for footnotes;
%% use the fntext command for theassociated footnote;
%% use the corref command within \author for corresponding author footnotes;
%% use the cortext command for theassociated footnote;
%% use the ead command for the email address,
%% and the form \ead[url] for the home page:
%% \title{Title\tnoteref{label1}}
%% \tnotetext[label1]{}
%% \author{Name\corref{cor1}\fnref{label2}}
%% \ead{email address}
%% \ead[url]{home page}
%% \fntext[label2]{}
%% \cortext[cor1]{}
%% \address{Address\fnref{label3}}
%% \fntext[label3]{}

\noindent
\LARGE{Single photon detection with SiPMs irradiated up to 10$^{14}$~cm$^{-2}$ 1-MeV-equivalent neutron fluence}
%% use optional labels to link authors explicitly to addresses:
%% \author[label1,label2]{}
%% \address[label1]{}
%% \address[label2]{}
\vspace{1cm}
\normalsize

%\thanks{paolo.carniti@mib.infn.it}
%\thanks{claudio.gotti@mib.infn.it}

M.~Calvi$^{1,2}$, P.~Carniti$^{1,2}$, C.~Gotti$^{1,2}$, C.~Matteuzzi$^{1}$, G.~Pessina$^{1}$
\vspace{0.5cm}

\footnotesize
\noindent
$^1$INFN, Sezione di Milano Bicocca, Piazza della Scienza 3, Milano 20126, Italy\\
$^2$Universit\`a di Milano Bicocca, Dipartimento di Fisica G. Occhialini, Piazza della Scienza 3, Milano 20126, Italy
%\cortext[cor]{Corresponding author}

\normalsize

\begin{abstract}
Silicon photomultipliers (SiPM) are solid state light detectors with sensitivity to single photons.
Their use in high energy physics experiments, and in particular in ring imaging Cherenkov (RICH) detectors, is hindered by their poor tolerance to radiation.
At room temperature the large increase in dark count rate makes single photon detection practically impossible already at 10$^{11}$ cm$^{-2}$ 1-MeV-equivalent neutron fluence.
% The situation is improved by cooling, but remains challenging.
The neutron fluences foreseen by many subdetectors to be operated at the high luminosity LHC range up to 10$^{14}$ cm$^{-2}$ 1-MeV-equivalent.
In this paper we present the effects of such high neutron fluences on Hamamatsu and SensL SiPMs of different cell size.
The advantage of annealing at high temperature (up to 175 \degC) is discussed.
We demonstrate that, after annealing, operation at the single photon level with a SiPM irradiated up to 10$^{14}$ cm$^{-2}$ 1-MeV-equivalent neutron fluence is possible at cryogenic temperature (77 K) with a dark count rate below 1~kHz.
\end{abstract}

%\begin{keyword}
% keywords here, in the form: keyword \sep keyword
%SiPM \sep Radiation damage \sep Neutron irradiation \sep Annealing \sep Cryogenic temperature \sep Single photon detection
%% PACS codes here, in the form: \PACS code \sep code

%% MSC codes here, in the form: \MSC code \sep code
%% or \MSC[2008] code \sep code (2000 is the default)

%\end{keyword}

%\end{frontmatter}

%% main text
\section{Introduction}
\label{sec:intro}

Silicon photomultipliers (SiPMs) are solid state light detectors with sensitivity that extends down to the single photon level.
They consist of arrays of single photon avalanche diodes (SPADs), or ``cells'', with typical dimensions in the tens of $\mu$m, connected in parallel, each SPAD with its own quenching resistor \cite{SiPMreview1, SiPMinvented}.
Being solid state devices, SiPMs are smaller and more robust than vacuum-based photomultiplier tubes (PMTs), and their photon detection efficiency can exceed that of PMTs.
They are biased at much lower voltage (tens of V) and they are insensitive to magnetic field.

The cell size is selected according to requirements on gain, photon detection efficiency (PDE) and dynamic range.
For the best sensitivity to a small number of photons, a larger cell size is preferable, increasing fill factor and hence PDE.
The signal generated by the avalanche in a SPAD depends on its capacitance and overvoltage, that is the bias voltage above breakdown.
At the same time, a large overvoltage increases secondary spurious events (crosstalk and afterpulses).
The best compromise is usually found at 2 V to 5 V above breakdown, depending on the specific SiPM model and on the application requirements.
Since the SPADs in a SiPM are very well matched, they have essentially identical gain, and the response of the device is linear with the number of photons (for low cell occupancy).
A SiPM can then clearly resolve the number of triggered SPADs, and therefore count the number of photons hitting the device at a given time.

The main drawback of SiPMs, compared to PMTs, is the high dark count rate (DCR).
At room temperature, the DCR is typically above 10~kHz/mm$^2$, about three orders of magnitude larger than PMTs.
The DCR depends strongly on temperature, and can be  reduced by a factor 2-3 for each 10 \degC~decrease.
Another major drawback that limits the use of SiPMs in the field of high energy physics is their sensitivity to radiation, in particular to displacement damage, as expected from devices whose operation depends on the properties of bulk silicon \cite{SiradReview1, SiradReview2}.
The most evident effect of radiation is an increase of DCR by several orders of magnitude, while PDE, gain and signal shape are generally unaffected \cite{SiPMrad1, SiPMrad11, SiPMrad2, SiPMrad20, SiPMrad21, SiPMrad22, SiPMrad3, SiPMrad31, SiPMrad32, SiPMrad4}.
Displacement of silicon atoms from the lattice caused by high energy hadrons creates clusters of defects (vacancies and interstitials) with energy levels in the band gap that favor the creation and recombination of electron-hole pairs in the depletion region of the device, increasing the DCR.
Degradation induced by displacement damage is noticeable starting at 10$^8$ cm$^{-2}$ 1-MeV-equivalent neutron fluence, and becomes critical at higher levels.

%This effect can be at least in part mitigated by annealing and by lowering the operating temperature.

\begin{table}[t]
\footnotesize
\begin{center}
\begin{tabular}{|r|l|l|l|}
\hline
Manufacturer & Model & Device size & Cell size \\
  \hline
  Hamamatsu & S13360-1325CS & 1.3 $\times$ 1.3 mm$^2$ & 25 $\mu$m\\
     & S13360-1350CS & 1.3 $\times$ 1.3 mm$^2$ & 50 $\mu$m\\
       & S13360-1375CS & 1.3 $\times$ 1.3 mm$^2$ & 75 $\mu$m\\
       \hline
       SensL & MicroFC-SMTPA-10020 & 1 $\times$ 1 mm$^2$ & 20 $\mu$m\\
    & MicroFC-SMTPA-10050 & 1 $\times$ 1 mm$^2$ & 50 $\mu$m\\
  \hline
\end{tabular}
\caption{Summary of SiPM models tested in this paper.}
\end{center}
\label{tab:SiPMs}
\end{table}

This paper will focus on the characterization of SiPMs irradiated up to 10$^{14}$ cm$^{-2}$ 1-MeV-equivalent neutron fluence, a level currently foreseen by several subdetectors to be operated at the high luminosity Large Hadron Collider (HL-LHC).
A summary of the tested models from Hamamatsu \cite{hamadsheet} and SensL \cite{sensdsheet} is given in table \ref{tab:SiPMs}.
Particular care was devoted to the possibility of detecting single photons, in view of a possible use in next generation ring imaging Cherenkov (RICH) detectors \cite{SiPMRICH1, SiPMRICH2}.
The DCR increase induced by radiation and its recovery by annealing were studied.
The behavior of irradiated SiPMs was also tested at cryogenic temperatures, down to 77~K.

\section{Neutron irradiation at LENA}

%Reattore, spettro del reattore, determinazione di fluenza equivalente attraverso ``NIEL hypothesis''.
%Di conseguenza, determinazione di potenza e durata degli irraggiamenti.
%Questo \cite{NeutronIrrad1} sembra essere lo standard da seguire.
%Questo un articolo che fa una misura simile con reattore TRIGA \cite{NEutronIrrad2}.
%Poi si cita Davide Chiesa?

%The devices were biased close to breakdown during irradiation, although displacement damage is not expected to depend on device bias or temperature.

The neutron irradiation was performed at the LENA laboratory (Pavia, Italy), in a TRIGA Mark II nuclear reactor~\cite{LENAreactor}.
Four sample holders were prepared, each consisting of a custom board, inserted in a plastic bullet, and sealed with black tape.
In each holder we placed one device for each SiPM type in table \ref{tab:SiPMs}.
The initial spread in breakdown voltage within each device type was about 300 mV.
Aside from this difference, the four holders could be considered identical before the irradiation.
%Devices of the same type were purchased from the same batch, and showed small initial spread in $V_{BR}$.
The bias voltage and the readout signals were provided by a 6 m long cable, which was also used to lower and extract the bullets from the reactor.
An automatized and remotely controllable test setup allowed continuous biasing and IV curve characterizations during the irradiation.

\begin{figure}
\centering
\includegraphics[width=0.7\linewidth]{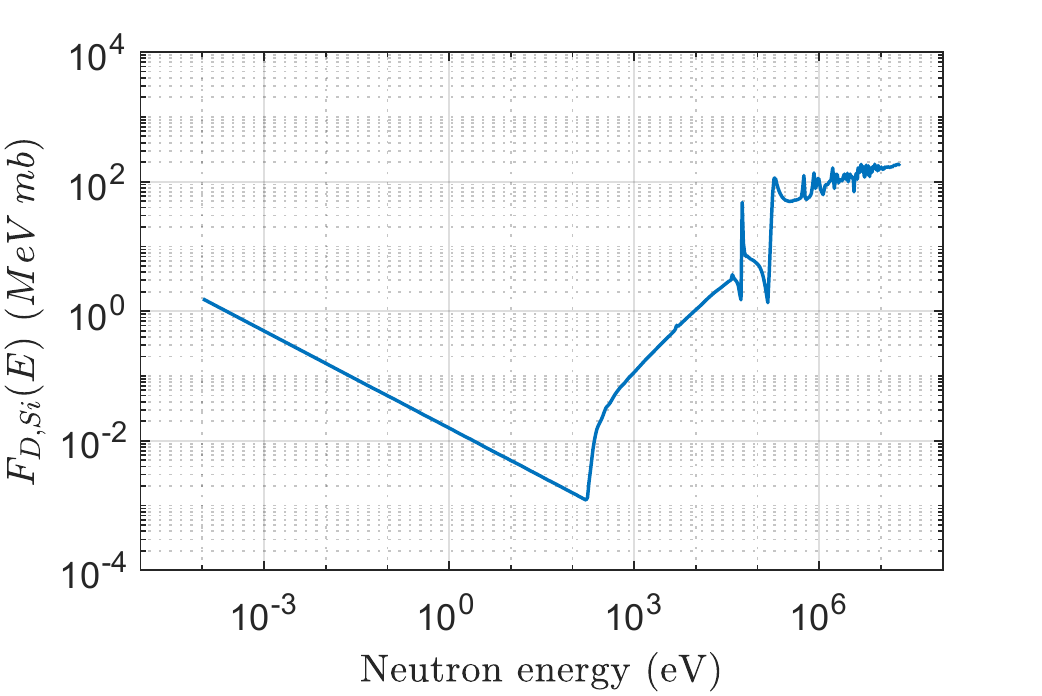}
\caption{Displacement damage function of silicon.}
\label{fig:irr_kerma}

\centering
\includegraphics[width=0.7\linewidth]{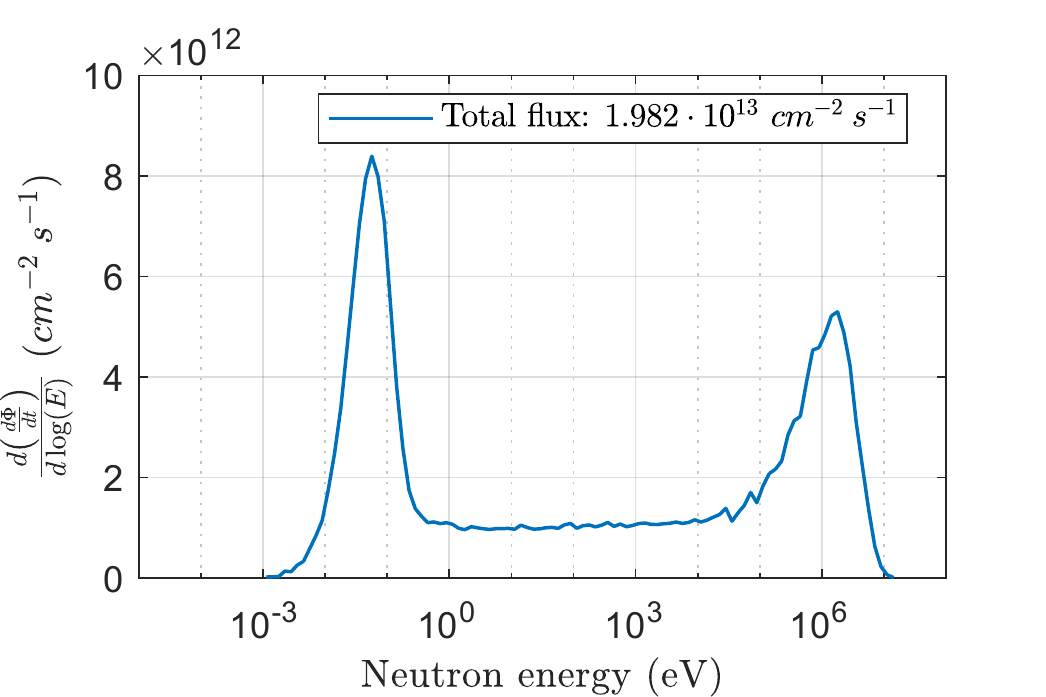}
\caption{Differential flux energy spectrum in the central thimble at the maximum reactor power (250~kW). The spectrum scales proportionally to reactor power.}
\label{fig:irr_spectrum}
\end{figure}

In neutron irradiation of silicon devices it is standard practice to express the irradiated fluence with the 1-MeV-equivalent neutron fluence in silicon  (n$_{eq}$/cm$^2$ in the following) \cite{NeutronIrrad1}.
The calculation of such parameter consists in an integral of the fluence energy spectrum, weighted by the displacement damage function in silicon (also known as KERMA function), shown in figure~\ref{fig:irr_kerma}.

In the LENA reactor, several channels are available for irradiation purposes, each one characterized by a different energy spectrum.
The channel that was chosen (central thimble) is the one with the hardest spectrum (higher fraction of fast neutrons) in order to match the maximum of the KERMA function.
Figure \ref{fig:irr_spectrum} shows the energy spectrum of the selected channel at the maximum reactor power (250~kW).
The data is taken from \cite{NeutronChiesa}.
Finally, the desired fluence was selected by choosing an appropriate reactor power and irradiation time.
The four chosen steps, $10^{11}$, $10^{12}$, $10^{13}$, and $10^{14}$ n$_{eq}$/cm$^{2}$, correspond to 47 minutes at reactor powers of 1.5, 15, 150 and 1500~W respectively.
Holder 1 was irradiated up to $10^{11}$ n$_{eq}$/cm$^{2}$, holder 2 was irradiated up to $10^{12}$ n$_{eq}$/cm$^{2}$, etc.

\section{Effects of displacement damage}
\label{sec:ivcurves}

\begin{figure}
\centering
\includegraphics[width=0.7\linewidth]{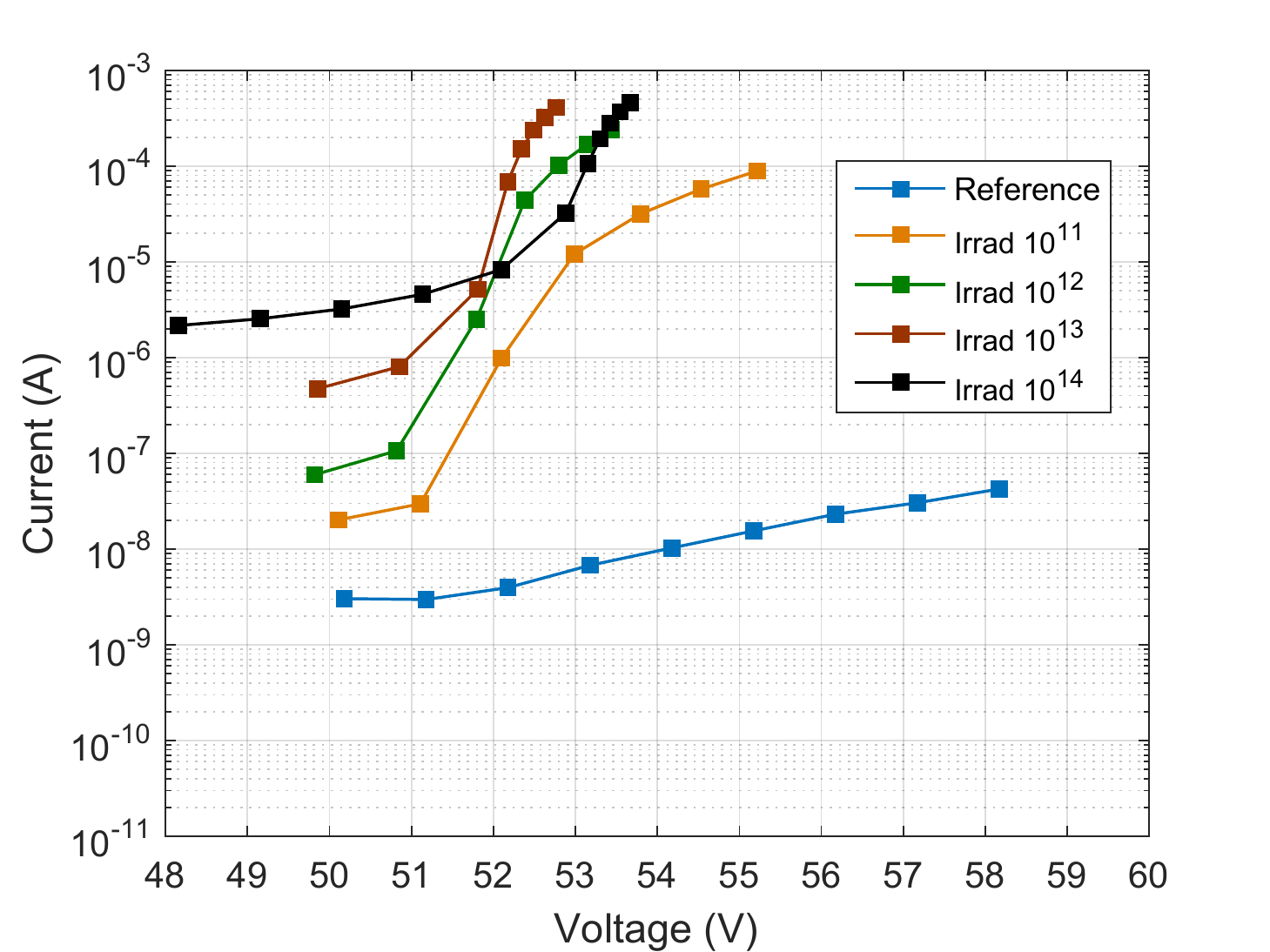}
\caption{Comparison of IV curves of Hamamatsu S13360-1325CS SiPMs at room temperature after irradiation, showing a large increase in current and a shift in breakdown voltage at 10$^{14}$ n$_{eq}$/cm$^2$.}
\label{fig:IVhama}

\centering
\includegraphics[width=0.7\linewidth]{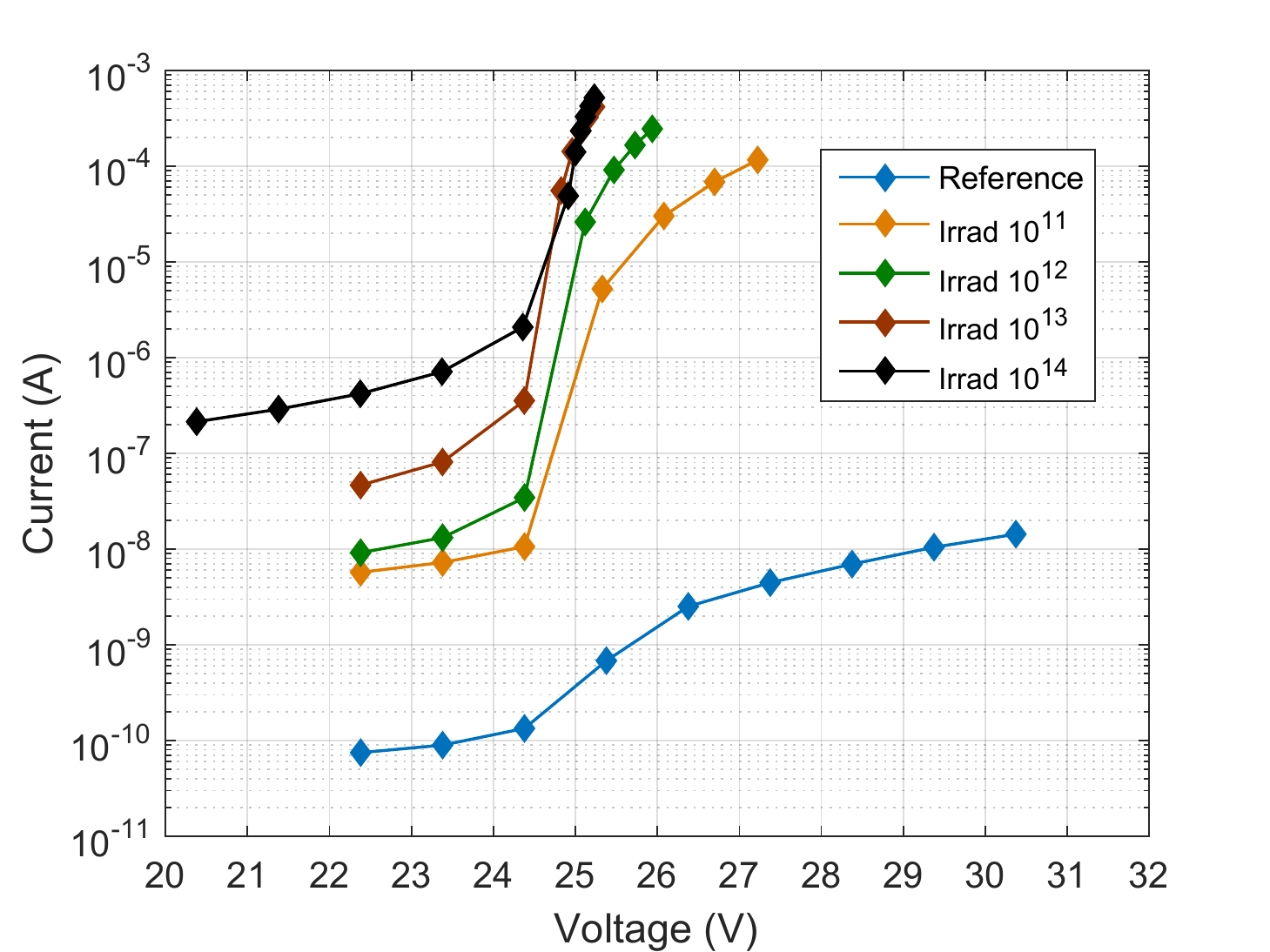}
\caption{Comparison of IV curves of SensL MicroFC-SMTPA-10020 SiPMs at room temperature after irradiation, showing a large increase in current but no significant shift in breakdown voltage up to 10$^{14}$ n$_{eq}$/cm$^2$.}
\label{fig:IVsens}
\end{figure}

Figures \ref{fig:IVhama} and \ref{fig:IVsens} show the IV curves of two SiPM models from Hamamatsu and SensL with similar cell size, irradiated at 10$^{11}$, 10$^{12}$, 10$^{13}$ and 10$^{14}$ n$_{eq}$/cm$^2$.
The curves were taken a few seconds after the end of the irradiation of each sample holder.
%For each plot, the curves shown belong to different samples, but their initial mismatch was negligible on this scale.
The current was measured across a 10~k$\Omega$ resistor in series with the devices.
This gives a sensitivity down to the pA range for unirradiated devices, and limits the current to fractions of mA for heavily irradiated devices.
The IV curves were performed by setting steps of 1 V before such resistor, while the voltage was measured after the resistor, directly across the SiPM.
As indicated in section 2, the initial spread in breakdown voltage between devices of the same type in different holders was about 300 mV; on this scale, the IV curves from different devices of the same type can therefore be compared directly.

\begin{figure}
\centering
\includegraphics[width=0.7\linewidth]{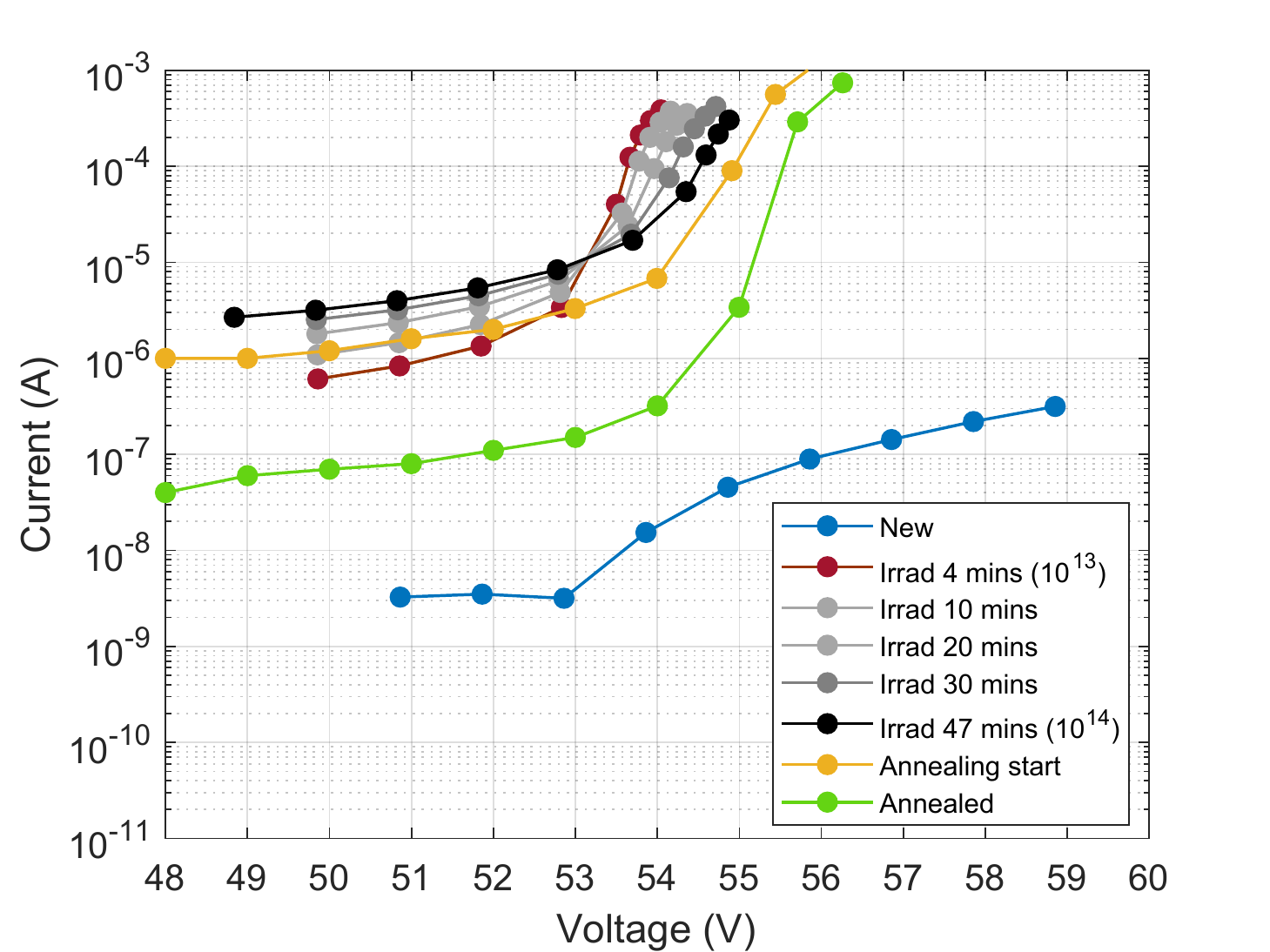}
\caption{IV curves of a Hamamatsu S13360-1350CS SiPM at room temperature before, during and after irradiation up to 10$^{14}$ n$_{eq}$/cm$^2$, which is reached after 47 minutes.
The curve taken after 4 minutes of irradiation corresponds to approximately 10$^{13}$ n$_{eq}$/cm$^2$. For fluence levels at the intermediate points during irradiation, see figure \ref{fig:IVhama50thr}.}
\label{fig:IVhama50}

\centering
\includegraphics[width=0.7\linewidth]{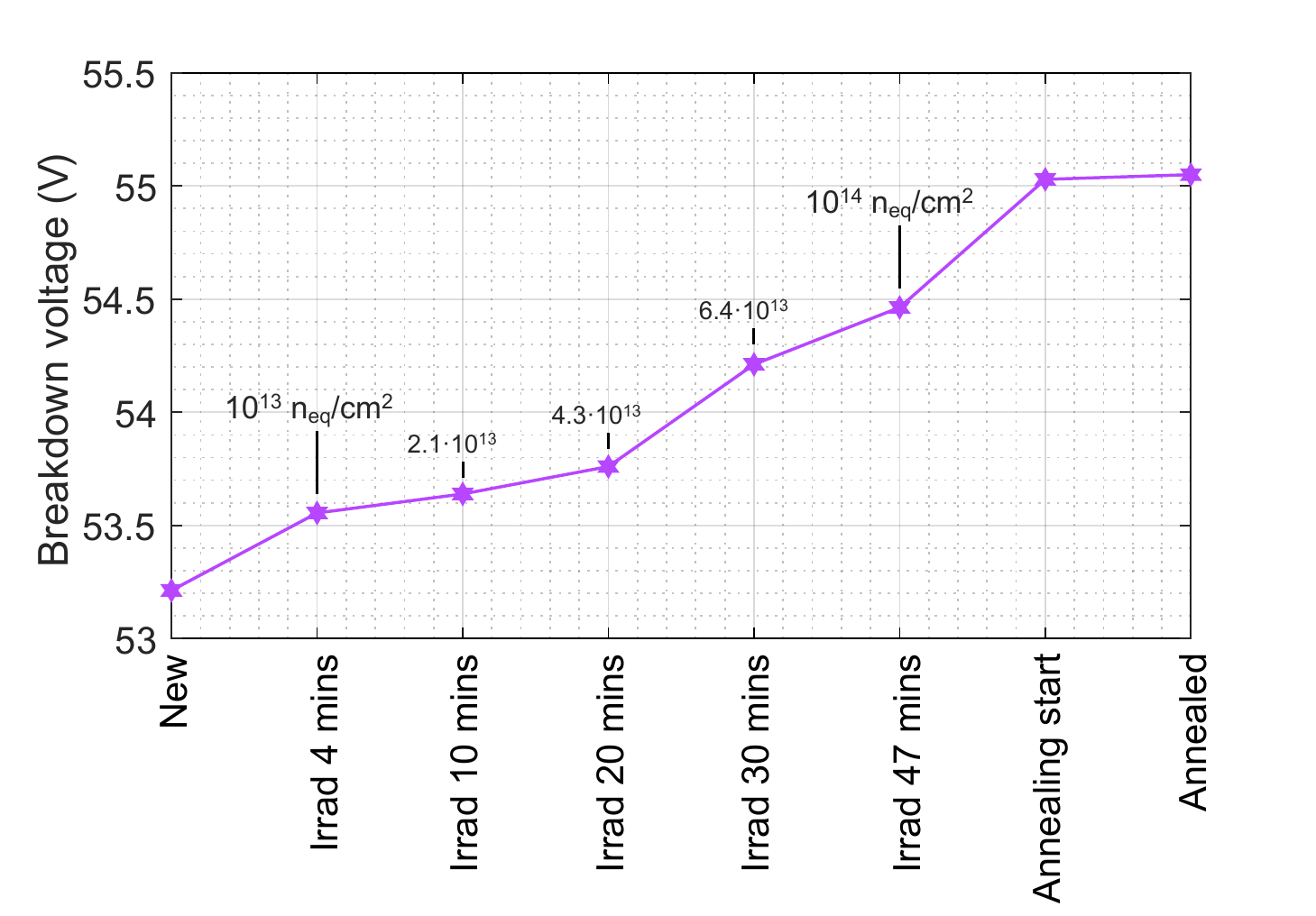}
\caption{Breakdown voltage extracted from the curves of figure \ref{fig:IVhama50}. The approximate fluence levels reached during irradiation are marked above the curve.}
\label{fig:IVhama50thr}
\end{figure}

A considerable degradation is already evident for both SiPM models at 10$^{11}$~n$_{eq}$/cm$^{2}$, with an increase in reverse current of 1-2 orders of magnitude below breakdown, and 3-4 orders of magnitude above breakdown.
Up to 10$^{13}$~n$_{eq}$/cm$^{2}$ the curves of the Hamamatsu devices differ only by the amount of leakage current, which is roughly proportional to the fluence both below and above breakdown.
Their breakdown voltage is unchanged up to 10$^{13}$~n$_{eq}$/cm$^{2}$, at least to the sensitivity level offered by this measurement, but shows a shift at 10$^{14}$~n$_{eq}$/cm$^{2}$.
The same effect was observed also in the other Hamamatsu devices of different cell size.
A possible interpretation of this shift might be related with the concentration of defects reaching large enough concentrations to interfere with the doping profiles of the device, modifying the shape of the electric field in the junction, hence the breakdown voltage.
There is however no way of verifying this hypothesis without detailed knowledge of the device structure, which is proprietary information of the manufacturer.
%As such, this intepretation remains speculative.
%This can be interpreted as follows: while at lower fluence the defects only enhance thermal generation and recombination of carriers, at 10$^{14}$~n$_{eq}$/cm$^{2}$ their concentration becomes large enough to interfere with the doping profiles of the device, modifying the shape of the electric field in the junction, hence the breakdown voltage.
This observation is at least qualitatively compatible with that reported in \cite{SiPMrad4}, although for a different device and manufacturer.

This effect is not observed up to 10$^{14}$~n$_{eq}$/cm$^{2}$ in the SensL devices, as shown in figure \ref{fig:IVsens}.
The difference in breakdown voltage between all the curves is smaller than 300~mV, and is compatible with the spread between different samples.
The SensL show a lower leakage current before irradiation, and also a smaller leakage current below breakdown in the case of the irradiated samples, by about one order of magnitude.
%This remains true even after scalinEven considering that their active area is smaller by a factor 1.7 compared to Hamamatsu
%Above breakdown, at the typical operating overvoltage, the situation is reversed, and the Hamamatsu devices exhibit a smaller current compared to SensL.
%Above breakdown, at the typical operating overvoltage, the situation is reversed, and the Hamamatsu devices exhibit a smaller current compared to SensL.
Above breakdown, at the typical operating overvoltage, the current of the irradiated samples is similar between Hamamatsu and SensL devices.

For each manufacturer, the IV curves for the other irradiated devices with different cell size are qualitatively very similar to those shown.
For larger cell size, signs of saturation are visible above the breakdown due to the smaller total number of cells.

\begin{figure}
\centering
\includegraphics[width=0.7\linewidth]{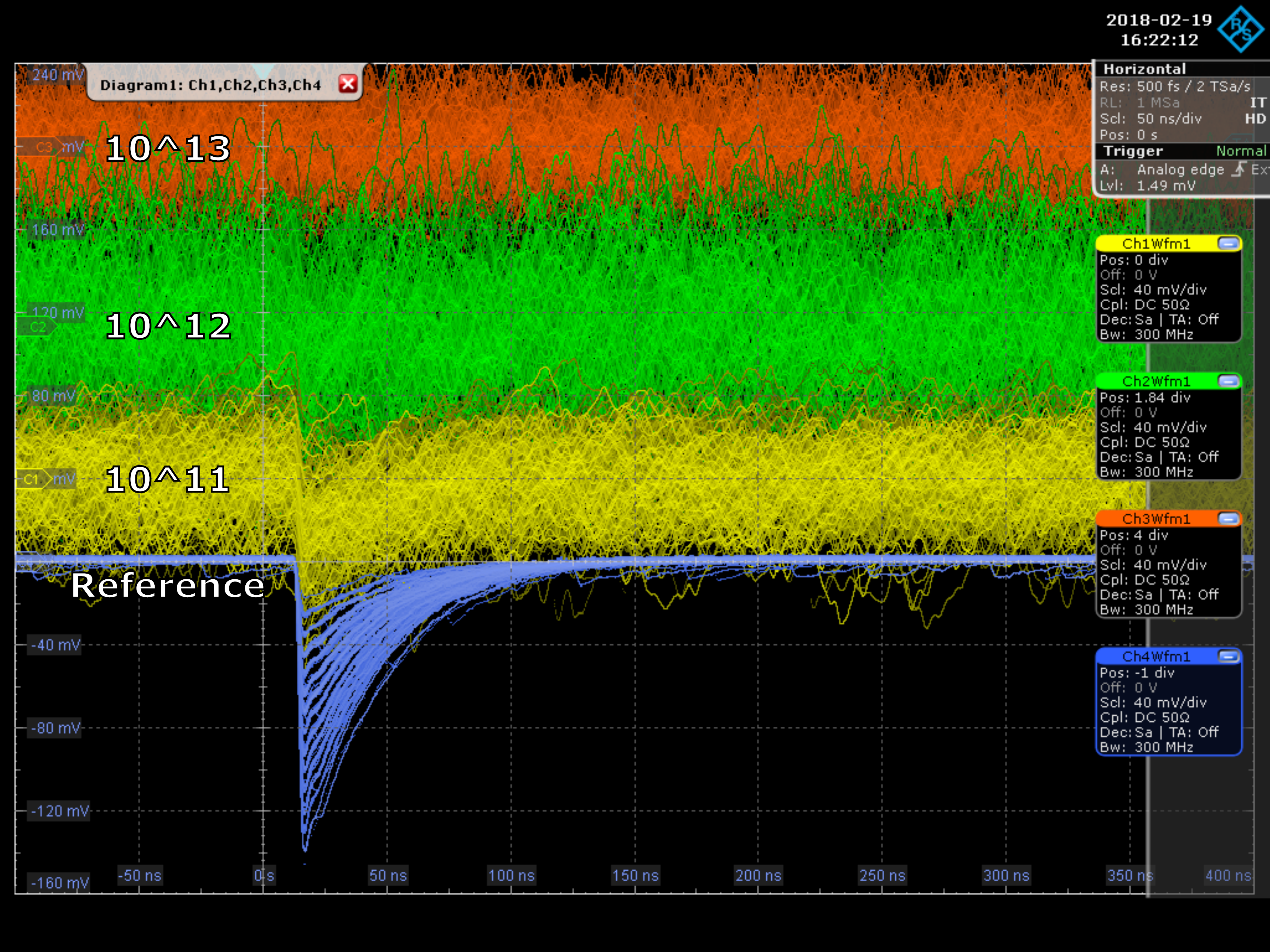}\\ \vspace{10px}
\includegraphics[width=0.7\linewidth]{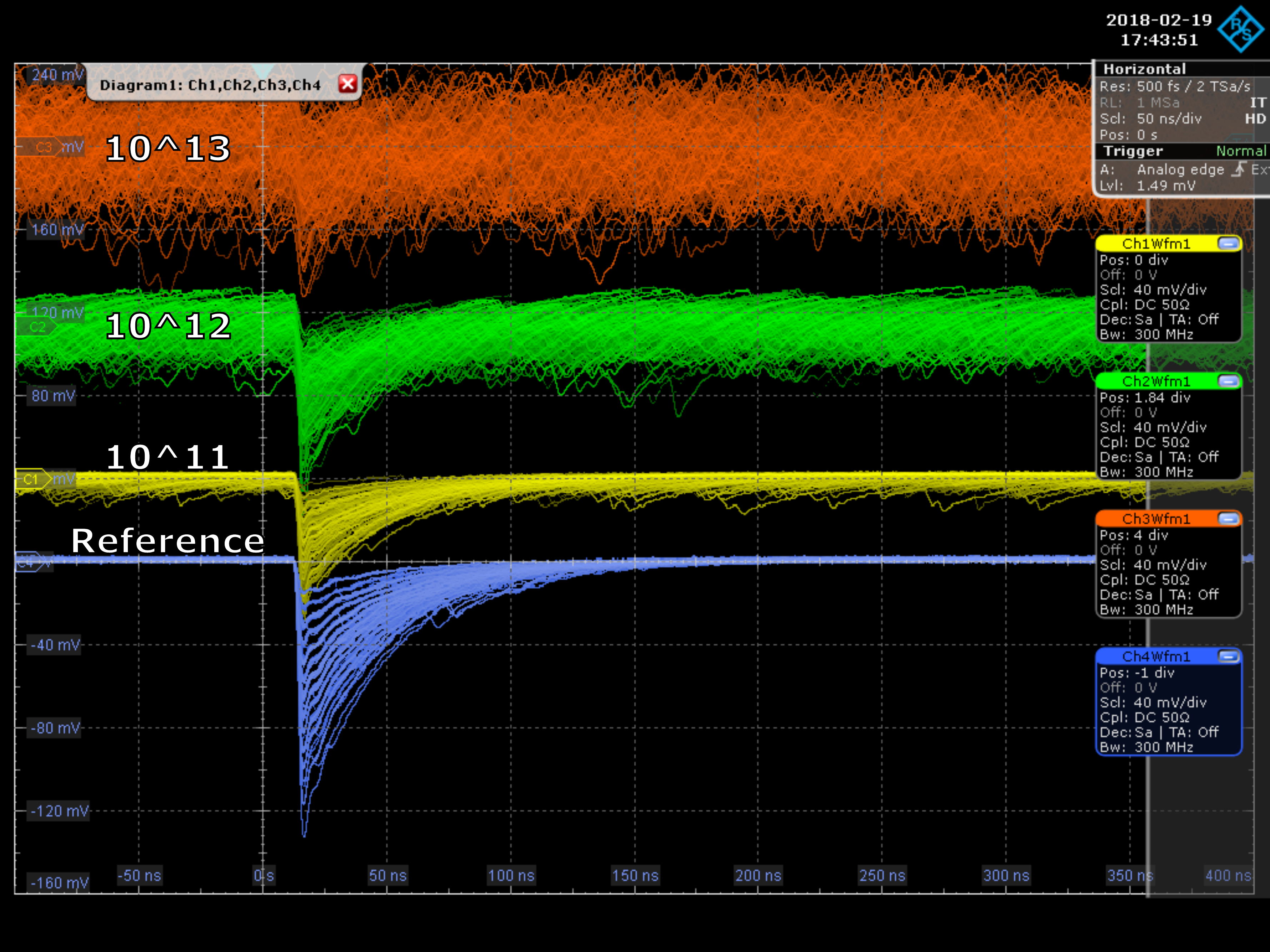}
\caption{Response to 400 nm laser pulses of a Hamamatsu S13360-1350CS SiPMs: non irradiated (blue) and irradiated up to 10$^{11}$ (yellow), 10$^{12}$ (green) and 10$^{13}$ (orange)  n$_{eq}$/cm$^{2}$. The measurement is performed at room temperature (upper image) and at $-30$ \degC~(lower image). The horizontal scale is 50 ns/div.
In this measurements all the SiPMs were biased at the same voltage. The reference device had a lower breakdown voltage than the others by about 0.7 V, which accounts for its higher gain.}
\label{fig:oscillolaser}
\end{figure}

The shift in breakdown voltage observed in the Hamamatsu devices can be examined in more detail in figure \ref{fig:IVhama50}, which shows the IV curves of a S13360-1350CS before, during and after irradiation up to 10$^{14}$ n$_{eq}$/cm$^{2}$.
The breakdown voltage was extracted from the IV curves with the method of the inverse logarithm derivative \cite{breakdownILD}.
The curves were interpolated with a polynomial, and the breakdown was calculated as the voltage where $\left(d \ln( I) / dV \right)^{-1}$ is minimum, and plotted in figure \ref{fig:IVhama50thr}.
%After 4 minutes of irradiation at 1500~W reactor power, the fluence is 10$^{13}$~n$_{eq}$/cm$^{2}$ and a breakdown voltage variation of about 300 mV is oserved.
The measurement shows a progressive shift of breakdown voltage during irradiation.
Assuming the shift is linear with fluence between 10$^{13}$ and 10$^{14}$ n$_{eq}$/cm$^{2}$, the precision of the extracted breakdown voltage can be estimated \textit{a posteriori} to be about 300 mV.
Compared to the curve taken before the irradiation (``new''), the breakdown shift amounts to about +1.3~V after 47 minutes in the reactor, when a fluence of 10$^{14}$~n$_{eq}$/cm$^{2}$ is reached.
The data marked as ``annealing start'' and ``annealed'' will be discussed in section \ref{sec:annealing}.

The irradiated devices were stored in a refrigerator at about 0 \degC~for several weeks after irradiation, before a deeper characterization could begin.
Comparison between the IV curves measured just after the irradiation and at the beginning of the characterization did not show significant differences.

Figure \ref{fig:oscillolaser} shows the response of the Hamamatsu S13360-1350CS to laser pulses, wavelength 400 nm, at room temperature and at $-30$ \degC.
The laser pulses are generated by a Hamamatsu PLP-10, and triggered by the oscilloscope (Rohde\&Schwarz RTO1044).
The signals are read out at the anodes by inverting amplifiers based on Texas Instruments LMH6702 current feedback opamps \cite{CFOA}.
The laser pulses contain a number of photons ranging between 4 and 15, as observed by the discrete signal amplitudes on the reference (non irradiated) sample.
All the SiPMs were biased with the same voltage.
The reference device in this measurement had a lower breakdown voltage than the others by about 0.7~V, which accounts for its higher gain.
In this measurement the devices were operated at lower overvoltage compared to the typical value of 3~V recommended by Hamamatsu. This does not necessarily represent the optimal working point of the devices, but the qualitative comparison between devices irradiated to different neutron fluences holds nonetheless.
At room temperature, due to the increased DCR, the pulses can hardly be seen on the sample irradiated to $10^{11}$ n$_{eq}$/cm$^{2}$, and are lost on the samples irradiated to higher levels.
At \mbox{$-30$ \degC}~the same laser pulses can still be seen, although barely, up to $10^{13}$ n$_{eq}$/cm$^{2}$, but could not be observed on the sample irradiated to $10^{14}$ n$_{eq}$/cm$^{2}$, not shown.
Even at such low temperature, operation at the single photon level is extremely challenging already at $10^{11}$ n$_{eq}$/cm$^{2}$.
Except for the increased DCR, no significant change to the gain or the signal shape could be noticed.

\begin{figure}
\centering
\includegraphics[width=0.7\linewidth]{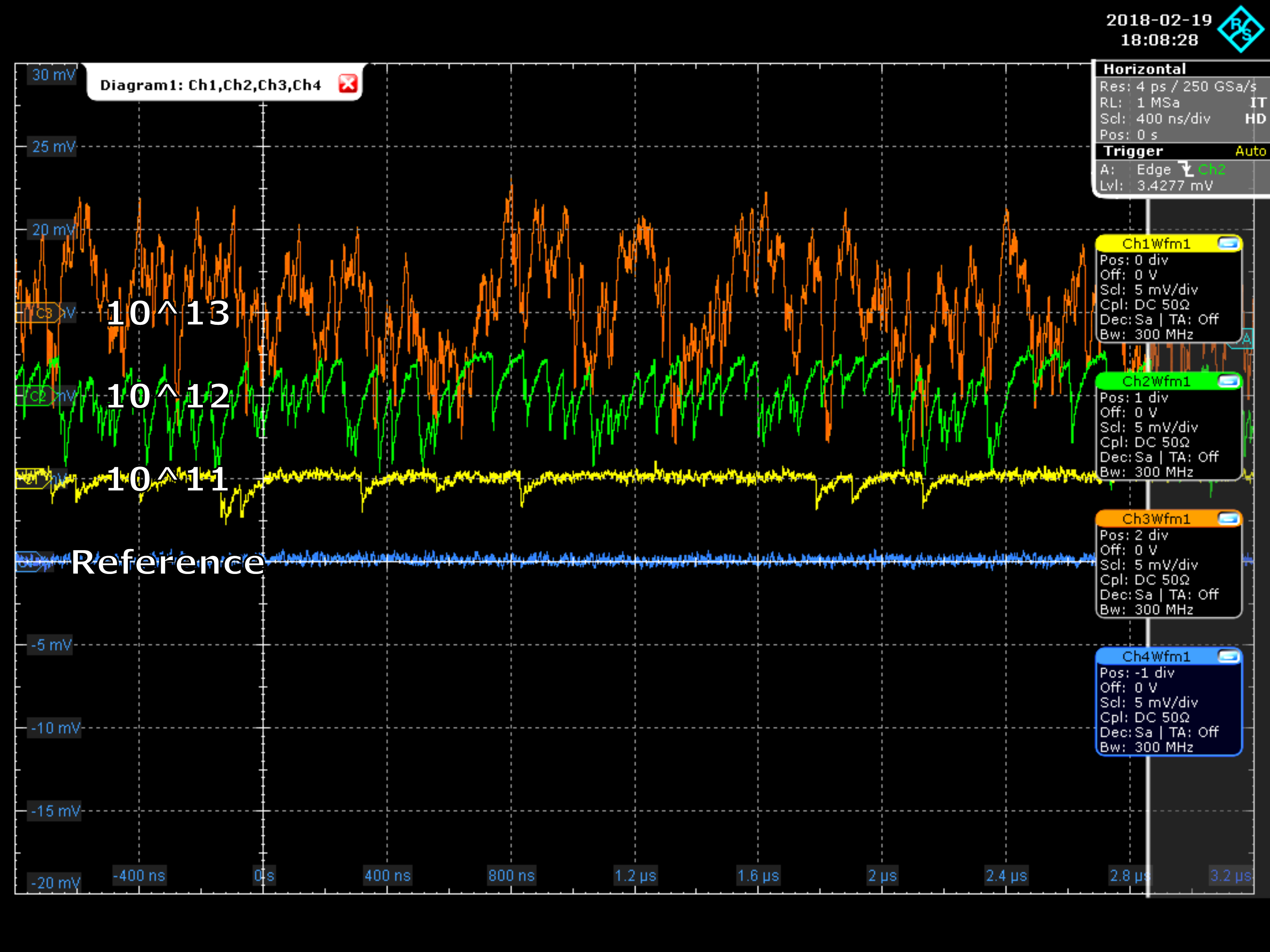}
\caption{Dark counts at $-30$ \degC \  of a Hamamatsu S13360-1350CS SiPMs biased at 51\ V: non irradiated (blue) and irradiated up to 10$^{11}$ (yellow), 10$^{12}$ (green) and 10$^{13}$ (orange) n$_{eq}$/cm$^{2}$. The horizontal scale is 400 ns/div.}
\label{fig:oscillodarkpre}
\end{figure}

Figure \ref{fig:oscillodarkpre} shows the baselines (dark counts only) of the same devices at $-30$ \degC.
The reference sample shows a dark count rate of about 500 Hz/mm$^2$.
The dark count rates of the irradiated samples are about 2 MHz/mm$^2$ at $10^{11}$ n$_{eq}$/cm$^2$ and 20 MHz/mm$^2$ at $10^{12}$ n$_{eq}$/cm$^2$.
It could not be clearly measured on the devices irradiated up to $10^{13}$ n$_{eq}$/cm$^2$, but a DCR of the order of 200 MHz/mm$^2$ can be inferred.

The reduction in DCR from room temperature to $-30$ \degC~is roughly a factor 100, or a factor 2.5 for each 10 \degC~ reduction in temperature in this temperature range.
This remains approximately valid for irradiated devices as well.
This indicates that the dominant source of DCR in this temperature range is of thermal origin, mediated by the concentration of traps in the band gap (Shockley-Read-Hall process), which linearly depends on the displacement damage dose.
Even at \mbox{$-30$ \degC}~, the DCR of irradiated devices is still critically high, and operation in the single photon range in these conditions appears extremely challenging, if not impossible.

\section{Recovery by annealing}
\label{sec:annealing}

After irradiation, the displaced atoms can diffuse back to their original positions in the lattice.
This annealing process can cure at least part of the defects created by radiation.
After several days at room temperature, a reduction in DCR of about a factor 2 has been reported \cite{SiPMrad1, SiPMrad2, SiPMrad3}.
For a given lattice defect type, by Arrhenius equation, the annealing rate is expected to show an exponential behavior with temperature.
As a very rough guess, it can be expected to double for every 10~\degC~increase in temperature.
Different classes of lattice defects are known to anneal at different temperatures \cite{SiradAnneal}.
Recovery of about one order of magnitude at room temperature and two orders of magnitude at 85~K  has been reported for SiPMs irradiated up to 10$^9$ n$_{eq}$/cm$^2$ and annealed at 250 \degC~ with forward bias applied \cite{SiPMAnneal}.

\begin{figure}
\centering
\includegraphics[width=0.7\linewidth]{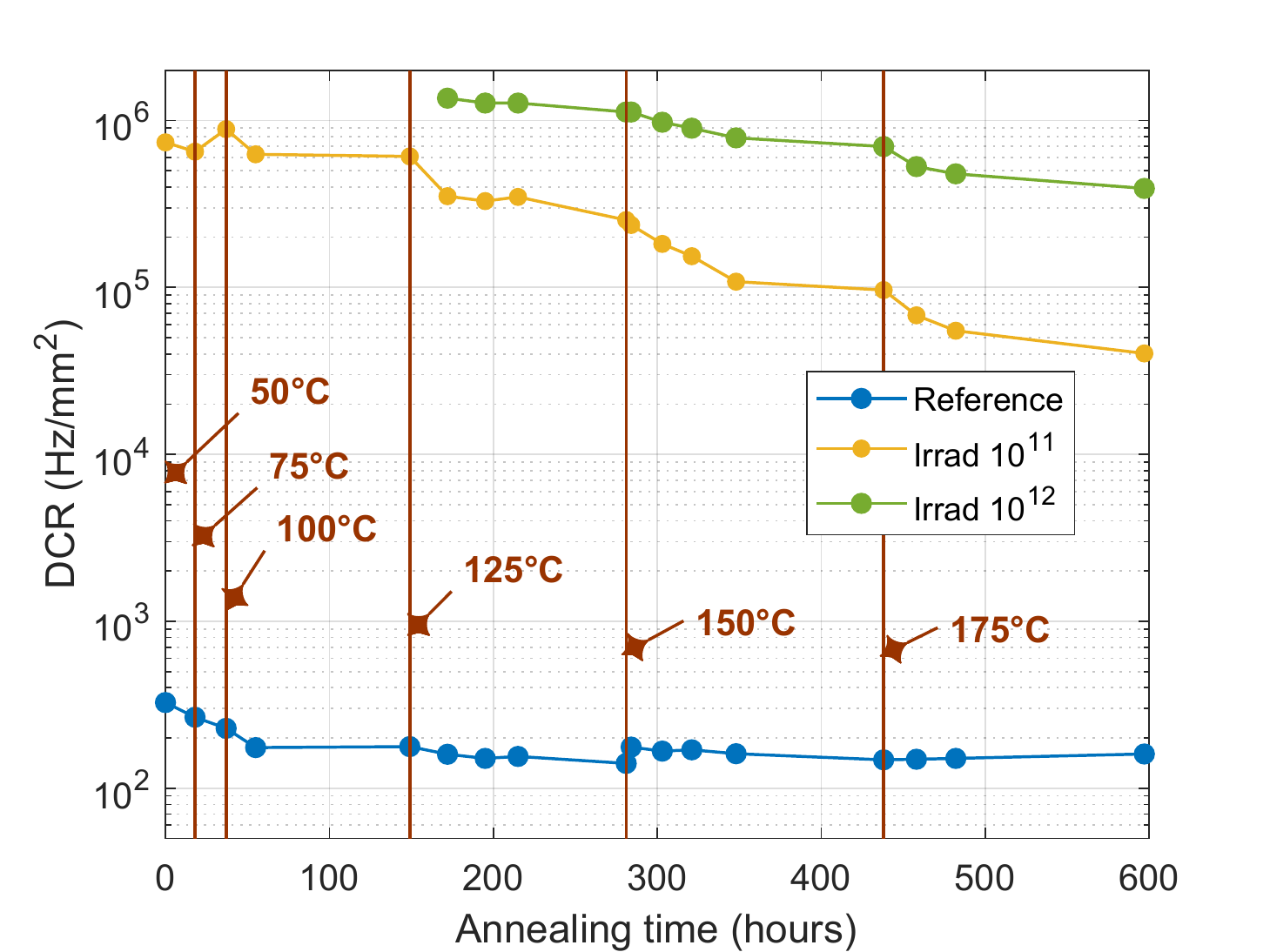}
\caption{Dark count rate (single photoelectron) at $-40$ \degC \ as a function of the annealing time. The annealing temperature is written close to the curve.}
\label{fig:annealing}
\end{figure}

To measure the annealing rate at high temperature, we placed the irradiated Hamamatsu S13360-1350CS SiPMs in a climatic chamber for a total time of 600 hours (about 1 month).
%Although applying forward bias to the device is reported in \cite{SiPMAnneal} to accelerate annealing,
Forward biasing of the SiPM junction \cite{SiPMAnneal} was not used in our measurements, in order to have better control of the annealing temperature; the use of longer annealing time and progressive temperature steps were preferred.
Several times during this period the SiPMs were extracted, cooled to \mbox{$-40$ \degC}~to measure the DCR, and then heated again to high temperature.
Figure \ref{fig:annealing} shows the measured DCR at  \mbox{$-40$ \degC}~as a function of the annealing time.
The annealing temperature was progressively increased from 50 \degC~up to 175 \degC, as indicated in the plot.
The measurement of the DCR at the single photoelectron level was possible only for the samples irradiated up to $10^{11}$ and $10^{12}$ n$_{eq}$/cm$^2$, as the maximum rate that could be detected in this measurement was about 2 MHz.
The DCR of a non irradiated sample, a few hundred Hz/mm$^2$, is shown for reference.
It is essentially constant, as expected, although a reduction in DCR of about a factor 2 can be noticed at the start of the curve.
A possible explanation is the presence of residual defects from device fabrication, which were cured in the first hours of the annealing period.

In the case of the sample irradiated up to 10$^{11}$  n$_{eq}$/cm$^2$, the measurement shows a reduction in DCR by nearly a factor 15, from above 700 kHz/mm$^2$ to about 40 kHz/mm$^2$.
A similar improvement is observed for the sample irradiated up to 10$^{12}$ n$_{eq}$/cm$^2$, from about 7 MHz/mm$^2$ (out of range for a precise determination) to 400 kHz/mm$^2$.
At each temperature step, the first 1-2 days give the largest reduction in DCR, with a constant slope in the log plot of figure \ref{fig:annealing} that is compatible with an exponential relaxation.
Longer annealing at the same temperature results in a smaller improvement, as the curve tends toward a plateau, before the temperature is increased again.
This is compatible with the presence of different classes of defects, with different annealing temperatures.

The IV curves of the device before and after annealing are shown in figure \ref{fig:IVhama50}.
The curve ``Annealing start'' is taken after several days of measurements at room temperature.
There are two significant differences with respect to the curve ``Irrad 47 mins'', taken shortly after the end of the irradiation to 10$^{14}$ n$_{eq}$/cm$^2$: the current is lower by about a factor 2 (compatible with what is expected from room temperature annealing \cite{SiPMrad1, SiPMrad2, SiPMrad3}), and the breakdown voltage is further shifted to the right by 500~mV, see figure \ref{fig:IVhama50thr}.
%The further change in breakdown voltage after the end of the irradiation might be related with the diffusion or drift at room temperarture of defects, that relocate in the hours and days after irradiation to find their final positions in the lattice.
The data marked as ``Annealed'' in figures \ref{fig:IVhama50} and \ref{fig:IVhama50thr} is taken after the annealing period at high temperature.
It shows a reduction of the current by about a factor 15 with respect to the previous curve.
This is consistent with the DCR reduction shown in figure \ref{fig:annealing} for the devices irradiated to lower fluence.
Notably, the breakdown voltage shift is identical to the previous curve, and is therefore not cured by annealing.
This might suggest that the defects responsible for breakdown shift are not affected by annealing, whereas those responsible for DCR and current increase are largely cured, at least up to the highest temperatures reached in this measurement (175~\degC).
However, it is stressed that no real understanding of the underlying mechanisms could be reached without dedicated studies, which would involve also a deep knowledge of the devide details.

\begin{figure}
\centering
\includegraphics[width=0.7\linewidth]{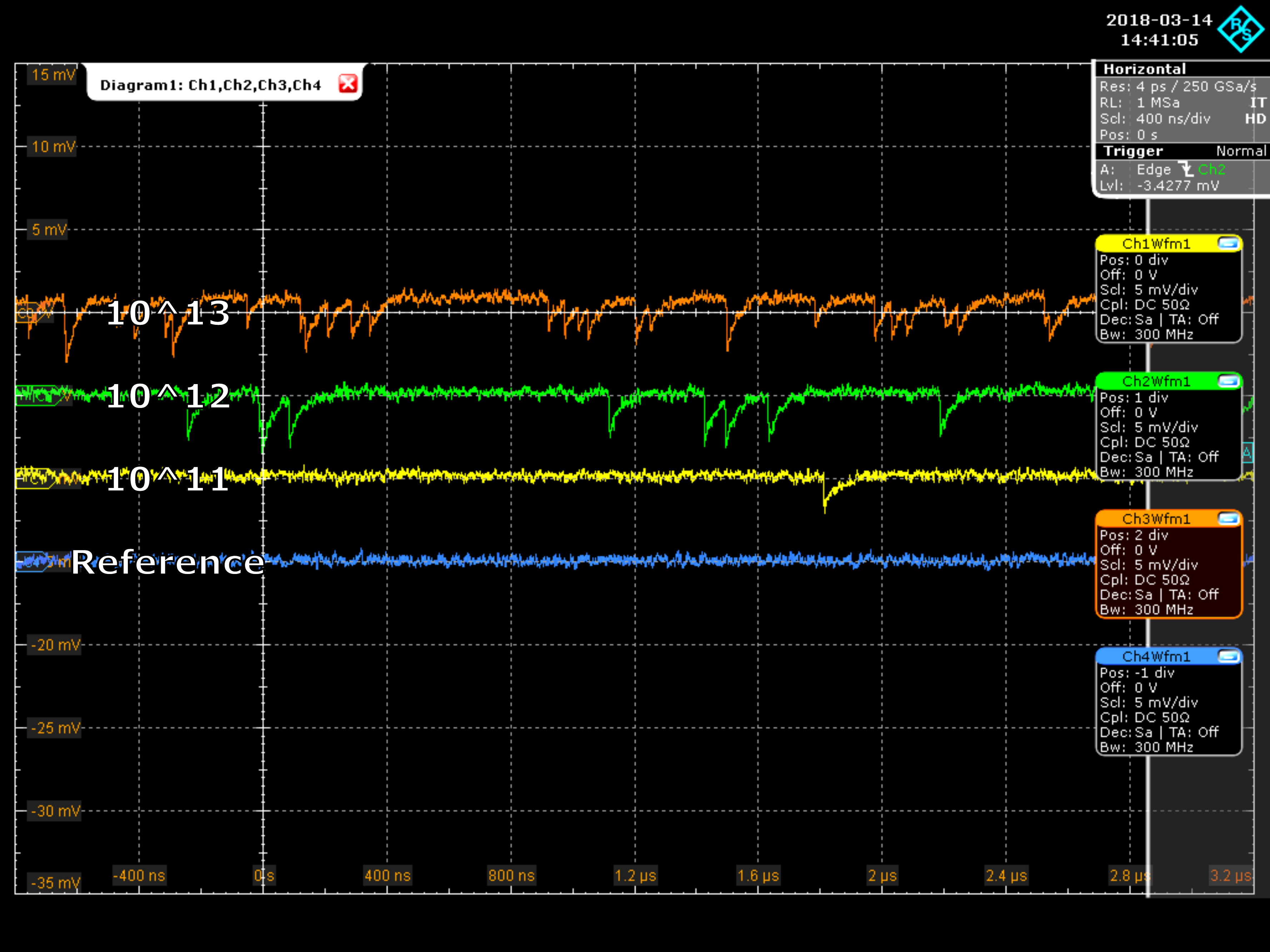}
\caption{Dark counts at $-30$ \degC, after the annealing process shown in figure \ref{fig:annealing}, of the same Hamamatsu S13360-1350CS SiPMs of figure \ref{fig:oscillodarkpre} biased at 51\ V: non irradiated (blue) and irradiated up to 10$^{11}$ (yellow), 10$^{12}$ (green) and 10$^{13}$ (orange) n$_{eq}$/cm$^{2}$. The horizontal scale is 400 ns/div.}
\label{fig:oscillodarkpost}
\end{figure}

Figure \ref{fig:oscillodarkpost}, taken at $-30$ \degC~after annealing, shows the baselines of the Hamamatsu S13360-1350CS in the same conditions as in figure \ref{fig:oscillodarkpre}.
The reduction in dark count rate is clearly evident by direct comparison between figures \ref{fig:oscillodarkpre} and  \ref{fig:oscillodarkpost}.
%A measurement of the IV curves, not shown, confirms a reduction of the leakage current of approximately the same factor.
%(about 15 for the device irradiated up to 10$^{11}$ n$_{eq}$/cm$^{2}$) both below and above breakdown.
%The shift in breakdown voltage observed for Hamamatsu devices at 10$^{14}$ n$_{eq}$/cm$^2$ is not recovered by annealing, although the current both below and above breakdown is reduced below the pre-annealing levels of the sample irradiated to 10$^{13}$ n$_{eq}$/cm$^2$.

Even though annealing at high temperature gives remarkable improvements in terms of DCR, it should be noted that such high temperatures could induce other changes in the devices, which were not fully assessed in the present measurement.
In particular, as reported by the manufacturer, long baking times above 100 \degC~could change the color of the resin that covers the device, resulting in a loss of sensitivity at short wavelength.
If annealing at high temperature is considered a viable method to extend the lifetime of these devices in radiation environments, this aspect should be studied, and possible alternatives to resin should be developed.

\section{Operation at cryogenic temperature}

As discussed in the previous sections, even at $-40$ \degC~and after annealing, single photon operation of irradiated SiPMs is still very challenging due to the high DCR.
In the range between room temperature and $-40$ \degC, the DCR is still dominated by the thermal generation of carriers. Further reduction in DCR is therefore expected by cooling the devices to lower temperature, below $-40$ \degC.
Operation at cryogenic temperatures was not trivial for older SiPM models, due to the large temperature coefficient of polysilicon quenching resistors \cite{SiPMreview1}.
Recent models, such as those we tested, have metal quenching resistors with small temperature coefficient, which allow to operate the SiPMs down to liquid nitrogen (77~K) with minimal impact on the signal shape \cite{SiPMAnneal}.

\begin{figure}
\centering
\includegraphics[width=0.7\linewidth]{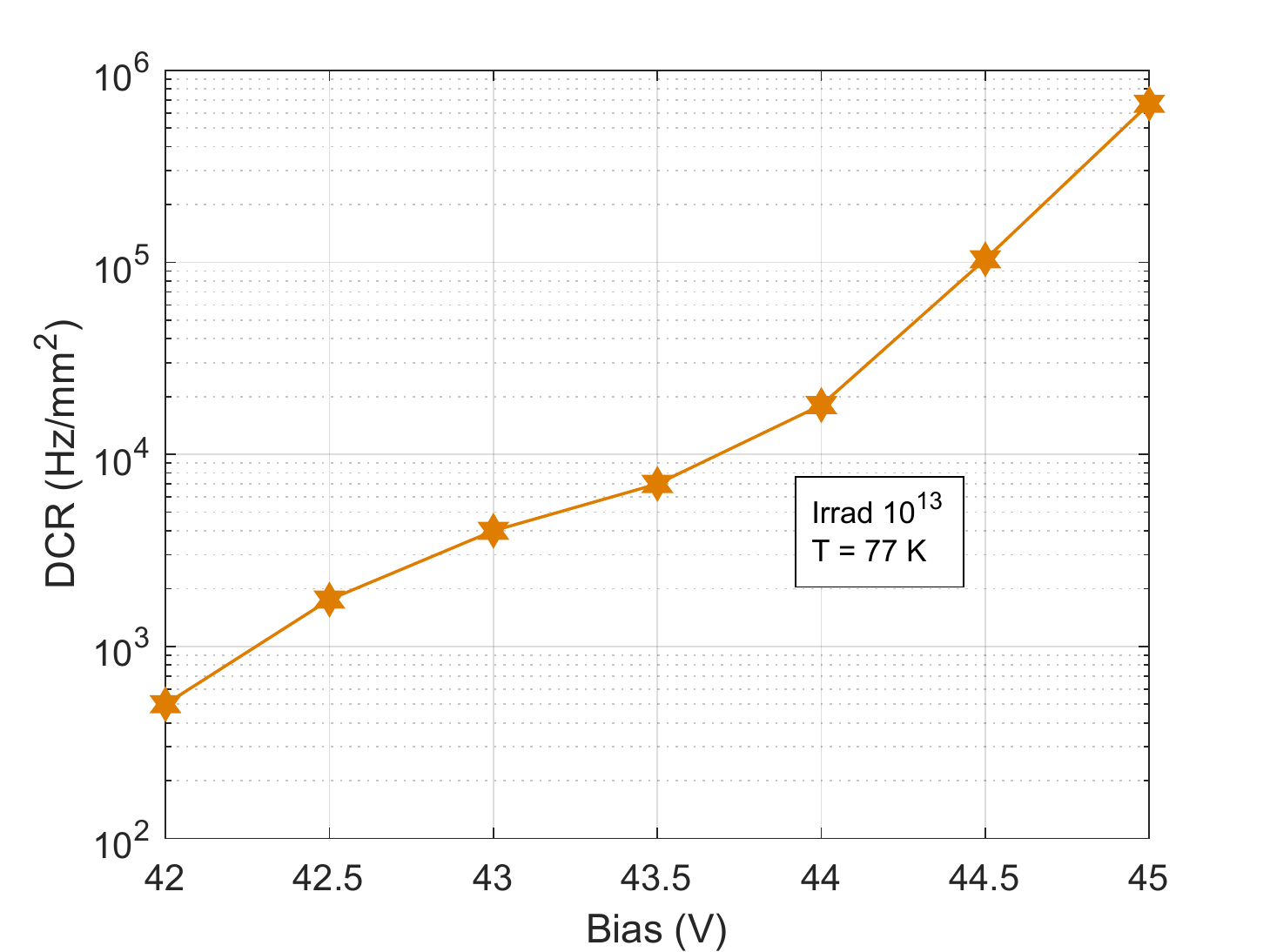}
\caption{DCR at 77~K versus bias voltage for a Hamamatsu S13360-1375CS device irradiated to 10$^{13}$ n$_{eq}$/cm$^2$.
The breakdown voltage of this device at 77~K is close to 40~V.
%The exponential dependence on bias voltage indicates tunnelling as the dominating mechanism of DCR.
}
\label{fig:tunnel}
\end{figure}

At liquid nitrogen temperature (77 K) the thermal generation of carriers is strongly suppressed, and the DCR is not of thermal origin anymore.
Unirradiated devices operated at this temperature show a DCR of the order of 1 Hz/mm$^2$, mainly contributed by band-to-band tunneling \cite{SiPMlowT}.
In the case of irradiated devices, the DCR is instead dominated by tunneling mediated by traps (trap-assisted tunneling), where the traps are the defects created by displacement.
In case of tunneling, a strong exponential dependence on the electric field, and therefore on overvoltage, is expected.
This is in contrast with the case of thermally generated DCR, that shows only a weak dependence on overvoltage.
As an example, figure \ref{fig:tunnel} shows the DCR measured at 77~K on a Hamamatsu S13360-1375CS device irradiated to 10$^{13}$ n$_{eq}$/cm$^2$, which spans three orders of magnitude between 2~V and 5~V overvoltage (the breakdown voltage of this device at 77~K is close to 40~V).
%The curve shows an exponential dependence on overvoltage, or on the electric field in the avalanche region, as is expected from tunneling.
This confirms trap-assisted tunnelling as the dominating mechanism of DCR at 77~K.
%, with the concentration of traps proportional to the total displacement dose.

\begin{figure}
\centering
\includegraphics[width=0.7\linewidth]{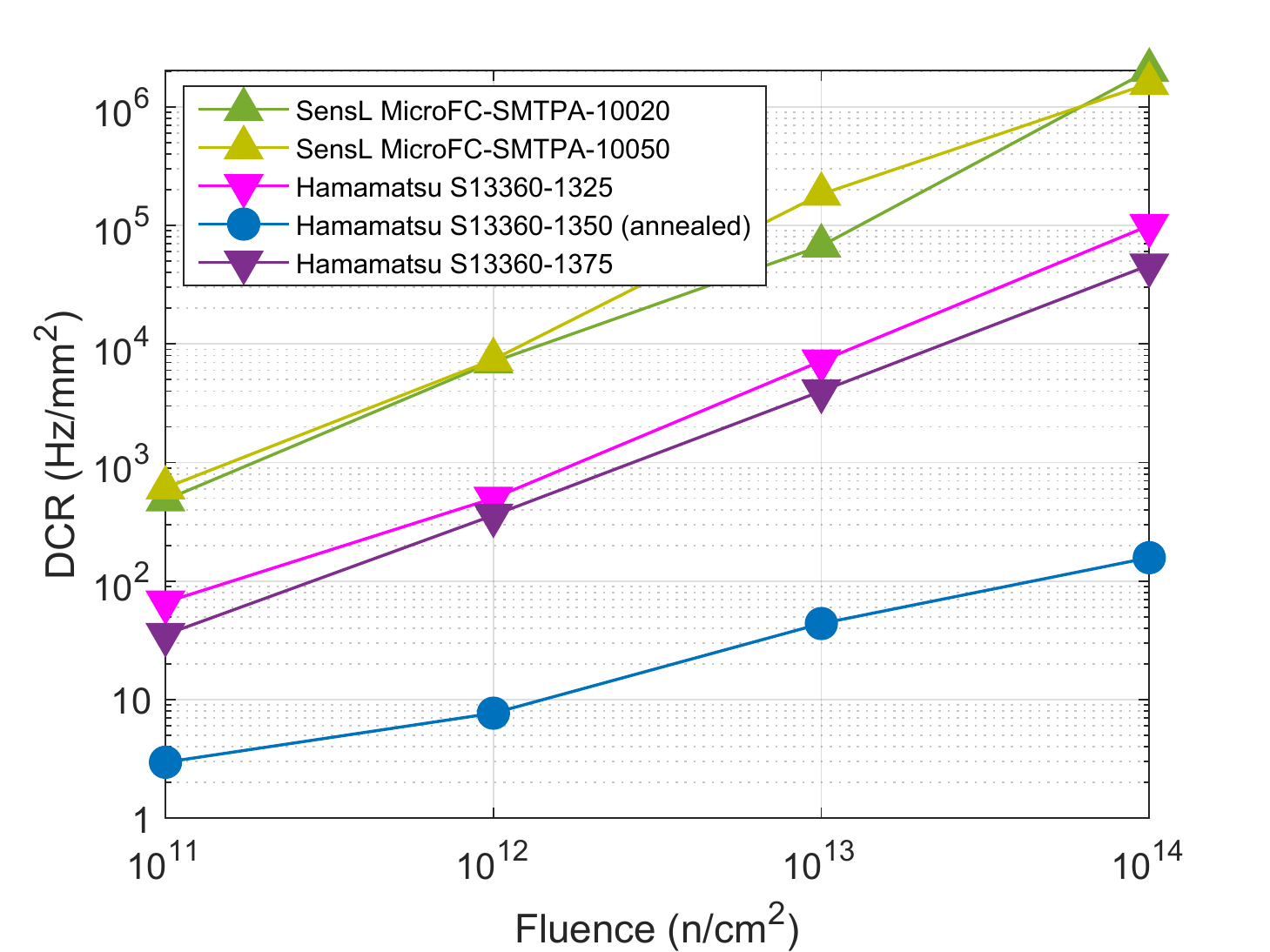}
\caption{DCR at 77\ K versus neutron fluence for all irradiated devices.
}
\label{fig:DCRvsirrad}
\end{figure}

The DCR of all the irradiated devices at 77~K is plotted in \mbox{figure} \ref{fig:DCRvsirrad} as a function of neutron fluence.
For this comparison, all the SiPMs were operated at the overvoltage recommended by the respective manufacturers at room temperature, that is 3~V for Hamamatsu S13360-1350CS and S13360-1375CS, 5~V for Hamamatsu  S13360-1325CS, and 2.5~V for the SensL devices.
The measurement shows that, as could be expected, the DCR of unannealed devices is in any case proportional to the neutron fluence.
The SensL devices show a higher DCR after irradiation compared to Hamamatsu devices.
At 10$^{14}$ n$_{eq}$/cm$^2$, the SensL devices show a DCR close to 1~MHz/mm$^2$, while the Hamamatsu devices are just below 100~kHz/mm$^2$.
%However, as was noted, in this temperature range the DCR shows an exponential dependence on overvoltage.
No significant dependence on the cell size was observed, at least when orders of magnitude are compared.
Smaller differences might be present, but are hard to disentangle from the exponential dependence of DCR on overvoltage.
%Although lowering the overvoltage reduces PDE and gain, the optimum operating point for a given application might be at a lower overvoltage than what manufacturers recommend at room temperature for unirradiated devices.
Lower DCR can be attained by lowering the overvoltage, although this reduces PDE and gain.
Further optimization might still be possible, to search for an operating point for SensL with a DCR closer to Hamamatsu.
A direct comparison between devices in these conditions is therefore not straightforward.

The Hamamatsu S13360-1350CS is the only device that went through annealing at high temperature, which explains the lower DCR compared to all the other devices.
The improvement in performance due to annealing is therefore confirmed also at cryogenic temperatures.
Also, the slope of the curve of the annealed device is smaller than the others, and breaks the proportionality with fluence.
This seems to indicate that the benefit of annealing observed at cryogenic temperature is actually higher at higher fluence levels.
%, from a factor 15 at 10$^{11}$ n$_{eq}$/cm$^2$ to a factor of about 150 at 10$^{14}$ n$_{eq}$/cm$^2$.
The DCR at 77~K of the annealed Hamamatsu device after 10$^{14}$ n$_{eq}$/cm$^2$ is well below 1~kHz/mm$^2$, and allows single photon detection.

\begin{figure}
\centering
\includegraphics[width=0.7\linewidth]{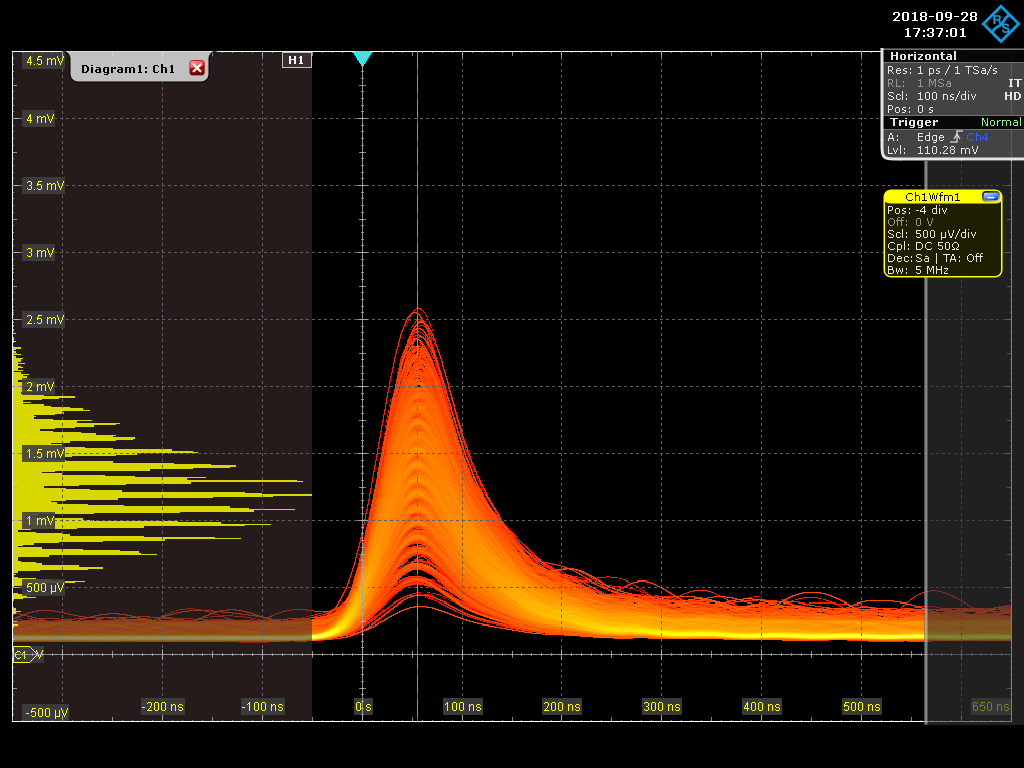}
\caption{Laser pulses on a Hamamatsu S13360-1350CS device, irradiated up to 10$^{14}$ n$_{eq}$/cm$^2$, annealed as described in section \ref{sec:annealing} and cooled to 77\ K. The horizontal scale is 100 ns/div.}
\label{fig:laser_LN2_irrad}
\end{figure}

\begin{figure}
\centering
\includegraphics[width=0.7\linewidth]{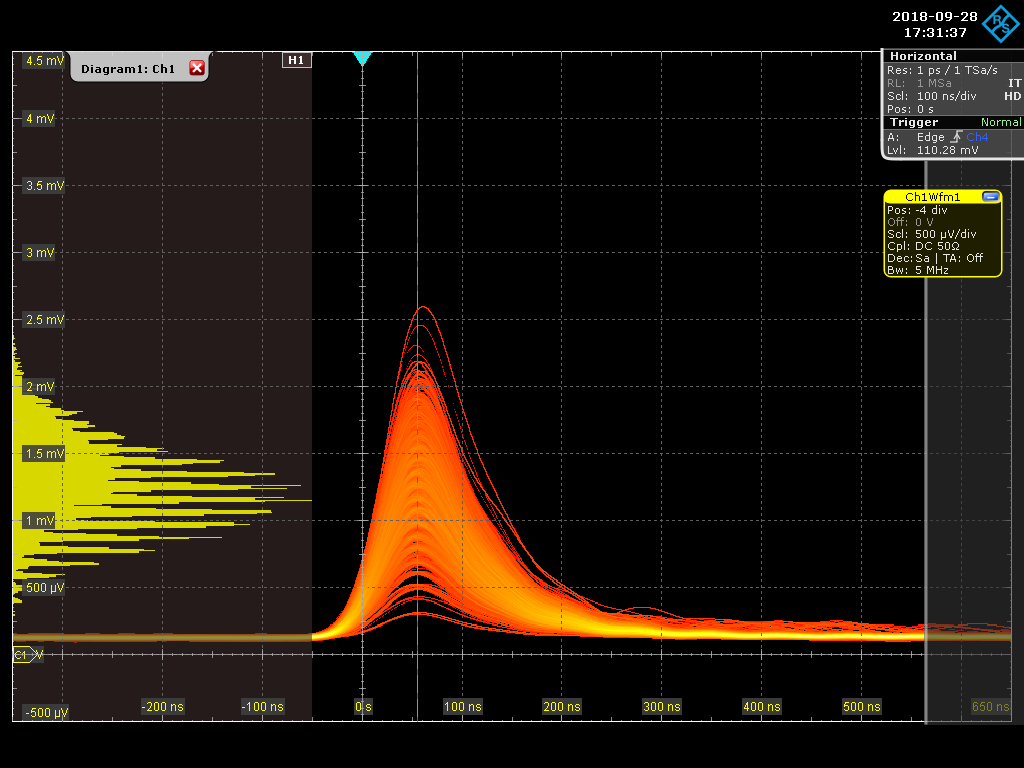}
\caption{Laser pulses on a new (unirradiated) Hamamatsu S13360-1350CS device cooled to 77\ K. The horizontal scale is 100 ns/div.}
\label{fig:laser_LN2_ref}
\end{figure}

Figure \ref{fig:laser_LN2_irrad} shows the response to 400 nm laser pulses at 77~K of a Hamamatsu S13360-1350CS SiPM, irradiated up to 10$^{14}$ n$_{eq}$/cm$^2$ and annealed as described in section \ref{sec:annealing}.
The laser repetition rate was 100~kHz, and the persistence of the oscilloscope display was set to 1~s.
The device was biased at 46.5\ V, corresponding to about 3 V overvoltage at 77~K, taking into account the breakdown voltage shift discussed in section \ref{sec:ivcurves}.
The laser pulses were propagated to the SiPM through an optical fiber.
The output signals were read out at the anode on a 50~$\Omega$ termination.
No amplifier was used, so the polarity is positive.
To eliminate noise from the environment, the bandwidth was limited to 5~MHz on the oscilloscope, which accounts for the longer rise time.
The amplitude histogram on the left side of the figure reveals an excellent separation of single photon signals.

%The response shows a reflection 50~ns after the signal due to sub-optimal cable connections in the setup.
%Once all this is considered, the signal shape is substantially unchanged with respect to room temperature.
% (compare with figure \ref{fig:oscillolaser}).
%Since the laser signals are now routed through an optical fiber, with a different setup than what was used in figure \ref{fig:oscillolaser}, the number of detected photons should not be compared directly between figures \ref{fig:laser_LN2} and \ref{fig:oscillolaser}, and only the signal shape and the DCR should be considered.
%Performing the same test with a non-irradiated device, lowering the operating temperature from room temperature to 77\ K gave no significant changes to the detection efficiency of the SiPM and to the transmissivity of the optical fiber.
The previous measurement can be compared with the same measurement obtained in the same conditions on a unirradiated (new) SiPM, shown in figure  \ref{fig:laser_LN2_ref}.
The SiPM was biased at 44~V, which again corresponds to about 3 V overvoltage at 77~K for this device, whose specified breakdown voltage at room temperature was 1~V lower than the specified breakdown voltage of the previous sample before irradiation and annealing.
When operated at this overvoltage, the gain of the SiPM is very close to the irradiated sample.
Particular care was taken in ensuring that the SiPM and the optical fiber had the same positions as in the previous measurement, with the goal of having the same amount of incident light.
In both measurements the maximum of the laser signal is at 10 photoelectrons, hence no variation of the PDE is observed for the irradiated and annealed device compared to the reference device.

Even after irradiation to 10$^{14}$  n$_{eq}$/cm$^2$, lowering the operating temperature is then effective in mitigating the increase in DCR, demonstrating operation down to the single photon regime.
Although for this study several SiPMs were repeatedly cooled to 77~K and no damage due to thermal stress was observed, this point should be carefully assessed if such low temperatures are to be used in a larger system, for instance in next generation RICH detectors.

%Stimare la temperatura a cui il tunnel diventa dominante rispetto agli eventi temrici (percui in sostanza non e' più vantaggioso scendere)?

\section{Conclusions}

The main effect of displacement damage on SiPMs is an increase in DCR, that makes single photon detection after 10$^{14}$ n$_{eq}$/cm$^2$ practically impossible at room temperature, or even at $-40$ \degC.
In this temperature range, noise is of thermal origin (Shockley-Read-Hall), mediated by the defect concentration, which is proportional to the total neutron fluence.
It shows a strong dependence on temperature, decreasing by about a factor 2.5 for each \mbox{$-10$ \degC}~variation, and a weak dependence on overvoltage.
A breakdown voltage shift of up to +1.8 V was observed in Hamamatsu S13360 series SiPMs after 10$^{14}$ n$_{eq}$/cm$^2$.
Annealing for a few days at high temperature, up to 175 \degC, was found to reduce the radiation-induced DCR by 1-2 orders of magnitude.
The breakdown voltage shift was not cured by annealing.
At liquid nitrogen temperature (77~K) the thermal noise is negligible, and the main source of DCR is tunneling mediated by defects (trap-assisted tunneling).
In this temperature range the DCR shows  weak dependence on temperature and a much stronger dependence on overvoltage, ranging by three orders of magnitude over the typical overvoltage range recommended by the manufacturers.
Single photon detection at 3 V overvoltage with a DCR below 1~kHz was possible with a SiPM irradiated up to 10$^{14}$ n$_{eq}$/cm$^2$, annealed for several days up to 175 \degC~and cooled to 77~K.

\section*{Acknowledgment}

The authors thank the staff at LENA, Pavia, Italy, for support during the irradiation of the samples.
%The authors also thank C.~D'Ambrosio for useful discussions.

%% The Appendices part is started with the command \appendix;
%% appendix sections are then done as normal sections
%% \appendix

%% \section{}
%% \label{}

%% If you have bibdatabase file and want bibtex to generate the
%% bibitems, please use
%%
%%  \bibliographystyle{elsarticle-harv} 
%%  \bibliography{<your bibdatabase>}

%% else use the following coding to input the bibitems directly in the
%% TeX file.

\end{document}